\documentclass[prc,superscriptaddress,noshowpacs,unsortedaddress,twocolumn,showpacs,preprintnumbers,amsmath,amssymb]{revtex4-1}

\usepackage{amsmath,amssymb,times}
\usepackage{color}
\usepackage{ulem}
\usepackage{bm}
\usepackage{here}

\usepackage{graphicx}

\def\lsim{~\,\makebox(1,1){$\stackrel{<}{\widetilde{}}$}\,~}

\newcommand{\beq}{\begin{equation}}
\newcommand{\eeq}{\end{equation}}
\newcommand{\bea}{\begin{eqnarray}}
\newcommand{\eea}{\end{eqnarray}}

\newcommand{\bfi}[1]{\mbox{\boldmath $#1$}}

\newcommand{\vK}{{\bfi K}}

\newcommand{\vs}{{\bfi s}}

\newcommand{\vrr}{{\bfi r}}
\newcommand{\vR}{{\bfi R}}

\def\a{\alpha}

\begin{document}
\title{ Neutron skin thickness of $^{208}$Pb, $^{116,120,124}$Sn, and $^{40}$Ca \\ 
determined from reaction cross sections of $^{4}$He scattering}

\author{Masayuki~Matsuzaki}
\affiliation{Department of Physics, Fukuoka University of Education,
Munakata, Fukuoka 811-4192, Japan}
%\email[]{matsuza@fukuoka-edu.ac.jp}

\author{Shingo~Tagami}
\affiliation{Department of Physics, Kyushu University, Fukuoka 819-0395, Japan}
%\email[]{sh.tagami@gmail.com}

\author{Masanobu Yahiro}
\email[]{orion093g@gmail.com}
\affiliation{Department of Physics, Kyushu University, Fukuoka 819-0395, Japan} 

\date{\today}

\begin{abstract}
\begin{description}
\item[Background]
We constructed the Kyushu chiral $g$-matrix and confirmed its reliability at 
$30  \lsim E_{\rm in} \lsim 100 $~MeV and $250  \lsim E_{\rm in} \lsim 400$~MeV 
for $^{12}$C scattering. 
Reaction cross section data of $^{4}$He scattering are available for some nuclides 
including $^{208}$Pb. 
PREX II collaboration reported a thick neutron skin for $^{208}$Pb. 
\item[Purpose]
Our purpose is to deduce neutron skin thicknesses of $^{208}$Pb and some other nuclides 
from reaction cross sections calculated in terms of the double folding model with 
the $g$-matrix. 
\item[Methods] 
We fold the $g$-matrix and densities given by mean field calculations. 
In order to remedy less-constrainedness of the  neutron sector, we renormalize densities 
so as to reproduce the observed cross sections. 
\item[Results] 
We found that a 3.4~$\%$ renormalization is necessary for $^{208}$Pb. 
The neutron density obtained from renormalization results in 
$R_{\rm skin}=$ 0.416$\pm$0.146 fm by confronting the precision proton radius. 
\item[Conclusions] 
Our result is consistent with PREX II
and therefore supports larger slope parameter $L$.
Results for $^{40}$Ca and $^{124}$Sn
are also consistent with $R_{\rm skin}$ deduced from other experiments.
For $^{116,120}$Sn the present method gives thicker skins.
\end{description}
\end{abstract}

\maketitle

%%%%%%%%%%%%%%%%%%%%%%%%%%%%%%%%%%%%%%%%%%%%%%%%%%%%%%%%%%%%%%%%%%%%%%%%%%%
%%%%%  Introduction 
%%%%%%%%%%%%%%%%%%%%%%%%%%%%%%%%%%%%%%%%%%%%%%%%%%%%%%%%%%%%%%%%%%%%%%%%%%%

\section{Introduction}
\label{Introduction}

In heavy atomic nuclei, neutrons outnumber protons as the mass number increases 
so as to mitigate the Coulomb repulsion between protons. 
This leads to the difference in the spatial distribution --- the neutron skin emerges, 
where the skin thickness is defined as the difference in the root mean square radii 
between neutrons and protons. 
This isovector property is not only one of the basic quantities in the structure of 
finite, terrestrial nuclei but determines the equation of state (EoS) of infinite nuclear 
matter in astrophysical objects such as neutron stars and exploding supernovae.

Information about nuclear radii is extracted from various experimental means. 
In contrast, theoretically, only mean-field calculations are available for heavy nuclei 
practically. 
Energy density functionals adopted in mean-field calculations contain many parameters 
of which numerical values are informed by basic observables such as binding energies, 
radii and so on, of representative stable and some unstable nuclides. 
The mean-field calculations predict various quantities including nuclear radii of 
other nuclides. 
Precision of proton radii among them is thought to be high because of cleanness of 
electron scatterings that inform the parameters of the proton sector. 
In contrast, neutron radii and accordingly skin thicknesses are less determined. 
This suggests that the calculated values of neutron radii should be critically assessed. 
In other words, it would be better to be based more directly on experimental information. 
One of such directions is to determine nuclear matter radius from reaction cross 
sections $\sigma_{\rm R}$ of nucleon-nucleus and/or nucleus-nucleus scatterings 
and then deduce neutron 
radius, and consequently skin thickness, by confronting the precision proton radius 
obtained by electron scatterings. 

As proposed by Horowitz et al.~\cite{PRC.63.025501}, on the other hand, parity-violating electron scatterings 
using polarized beams give directly neutron radii. By confronting them with the proton radii, 
skin thickness can be obtained. Actually, the PREX II experiment reported a precision datum 
of the neutron skin thickness of $^{208}$Pb~\cite{Adhikari:2021phr}; namely, 
\bea
R_{\rm skin}^{208}({\rm PREX~II}) = 0.283\pm 0.071\,{\rm fm}.
\eea 
The $R_{\rm skin}^{208}({\rm PREX~II})$ gives larger the slope 
parameter $L$ and supports stiffer EoSs. 
As a famous EoS, we can consider APR~\cite{Akmal:1998cf}. It yields 
$R_{\rm skin}^{208}=0.16$~fm. This value is out of $R_{\rm skin}^{208}({\rm PREX~II})$. 
This is an interesting issue to be solved, since this calculation is believed to be best for symmetric and neutron matter.   As for the density dependence of the symmetry energy, studied with heavy-ion collisions, collective excitation in nuclei (especially Pygmy Dipole Resonances) and neutron-star calculations, a good brief review is shown in 
Ref.~\cite{Li:2014oda}.

In relation to the present subject, our group has been studying nuclear reaction 
observables, including $\sigma_{\rm R}$ relevant to the present purpose, in 
terms of a microscopic optical potential based on a chiral $g$-matrix~\cite{Toyokawa:2017pdd}. This 
$g$-matrix was constructed by Kohno~\cite{PRC.88.064005} by taking into account the next-to-next-to-next-to 
leading order (N$^{3}$LO) two-body force and the NNLO three-body force in chiral 
perturbation. 
Toyokawa et al. localized the non-local $g$-matrix, and we call it the Kyushu chiral 
$g$-matrix~\cite{Toyokawa:2017pdd}. Its numerical values for selected discrete energies are presented in a 
web page for public use~\cite{localized}. 
In that work, $\sigma_{\rm R}$ of $^{4}$He + $^{58}$Ni and $^{4}$He + $^{208}$Pb 
were studied paying attention to the effect of the three-body force, in terms of the 
double-folding model (DFM) adopting a microscopic density of the Gogny-D1S Hartree-Fock 
(HF) for the targets and a phenomenological one~\cite{C12-density} for the projectile. 

In Ref.~\cite{Tagami:2019svt}, we first predicted the ground-state properties, such as binding energies, 
one- and two-neutron separation energies and various radii, of Ca isotopes adopting Gogny-
D1S Hartree-Fock-Boboliubov (HFB) with and without the angular-momentum projection (AMP). 
Using the nuclear densities given by this structure calculation and the Kyushu chiral $g$-matrix, 
we predicted $\sigma_{\rm R}$ for scattering of Ca isotopes on a $^{12}$C target 
with DFM, after confirming its reliability at each incident energy for 
$^{12}$C + $^{9}$Be, $^{12}$C, and $^{27}$Al scatterings. 

After the PREX II result~\cite{Adhikari:2021phr} is announced, we performed 
a single-folding model calculation of $\sigma_{\rm R}$ 
of p + $^{208}$Pb scattering by adopting the Kyushu chiral $g$-matrix~\cite{Tagami:2020bee} and the Gogny-HFB. 
The important finding of this study is that the calculated $\sigma_{\rm R}$ are 3~$\%$ 
smaller than the experimental values in the energy range in which the reliability of the 
Kyushu chiral $g$-matrix has been confirmed and the Gogny HFB reproduces the observed 
proton radii well. 
Then we assume that this originates from the less-confirmed mean-field parameters for 
the neutron sector and we attempted to renormalize the HFB+AMP neutron density to 
reproduce the $\sigma_{\rm R}$ data. 
The neutron radius deduced from the energy-averaged $\sigma_{\rm R}$ through the matter 
radius leads to a neutron skin thickness that agrees with the PREX II result well. 

The purpose of the present work is to examine further the present method --- extract 
the neutron radius from $\sigma_{\rm R}$ given by the Kyushu chiral $g$-matrix and the 
phenomenologically renormalized mean-field density --- by revisiting the $^{4}$He + $^{208}$Pb 
scattering studied in Ref.~\cite{Toyokawa:2017pdd} and comparing with the p + $^{208}$Pb result 
of Ref.~\cite{Tagami:2020bee}. 
Then we study some lighter nuclides. 

\section{Model}
\label{Sec-Framework}

The model adopted in this work is essentially the same as that in Ref.~\cite{Tagami:2020bee}, 
aside from the optical potential is obtained by double folding for $^{4}$He + $^{208}$Pb, 
rather than single folding for p + $^{208}$Pb. The double folding is performed 
for the Kyushu chiral $g$-matrix and the adopted nuclear densities. 

As the densities of $^{208}$Pb we newly examined the Skyrme HFB~\cite{Schunck:2016uvm} 
with the SLy7 parameter set, 
which is an improved version of the widely used SLy4~\cite{Chabanat:1997un}, in addition to the D1S-GHFB+AMP 
ones~\cite{Tagami:2019svt}. As for $^{4}$He, again we use the phenomenological 
density~\cite{C12-density}. 

The potential $U$ consists 
of the direct part ($U^{\rm DR}$) and the exchange part ($U^{\rm EX}$):
\bea
\label{eq:UD}
U^{\rm DR}(\vR) \hspace*{-0.15cm} &=& \hspace*{-0.15cm} 
\sum_{\mu,\nu}\int \rho^{\mu}_{\rm P}(\vrr_{\rm P}) 
            \rho^{\nu}_{\rm T}(\vrr_{\rm T})
            g^{\rm DR}_{\mu\nu}(s;\rho_{\mu\nu}) d \vrr_{\rm P} d \vrr_{\rm T}, \\
\label{eq:UEX}
U^{\rm EX}(\vR) \hspace*{-0.15cm} &=& \hspace*{-0.15cm}\sum_{\mu,\nu} 
\int \rho^{\mu}_{\rm P}(\vrr_{\rm P},\vrr_{\rm P}-\vs)
\rho^{\nu}_{\rm T}(\vrr_{\rm T},\vrr_{\rm T}+\vs) \nonumber \\
            &&~~\hspace*{-0.5cm}\times g^{\rm EX}_{\mu\nu}(s;\rho_{\mu\nu}) \exp{[-i\vK(\vR) \cdot \vs/M]}
            d \vrr_{\rm P} d \vrr_{\rm T},~~~~
            \label{U-EX}
\eea
where $\vs=\vrr_{\rm P}-\vrr_{\rm T}+\vR$ 
for the coordinate $\vR$ between the projectile (P) and target (T). The coordinate 
$\vrr_{\rm P}$ 
($\vrr_{\rm T}$) denotes the location for the interacting nucleon 
measured from the center-of-mass of P (T). 
Each of $\mu$ and $\nu$ stands for the $z$-component
of isospin; 1/2 means neutron and $-$1/2 does proton.
The original form of $U^{\rm EX}$ is a non-local function of $\vR$,
but  it has been localized in Eq.~\eqref{U-EX}
with the local semi-classical approximation in which
P is assumed to propagate as a plane wave with
the local momentum $\hbar \vK(\vR)$ within a short range of the 
nucleon-nucleon interaction, where $M=A A_{\rm T}/(A +A_{\rm T})$
for the mass number $A$ ($A_{\rm T}$) of P (T).
The validity of this localization is shown in Ref.~\cite{Minomo:2009ds}.

The direct and exchange parts, $g^{\rm DR}_{\mu\nu}$ and 
$g^{\rm EX}_{\mu\nu}$, of the effective nucleon-nucleon interaction 
($g$-matrix) depend on the local density
\bea
 \rho_{\mu\nu}=\sigma^{\mu} \rho^{\nu}_{\rm T}(\vrr_{\rm T}+\vs/2)
\label{local-density approximation}
\eea
at the midpoint of the interacting nucleon pair, where $\sigma^{\mu}$ is the Pauli matrix of a nucleon in P. 
This choice of  the local density is quite successful for $^{4}$He scattering, as shown in Ref. \cite{PRC.89.064611}. 

The renormalization, that is, the scaling of the density $\rho(\vrr)$, will be performed 
as follows: 
We can obtain the scaled density $\rho_{\rm scaling}(\vrr)$ from the original density $\rho(\vrr)$ as
\bea
\rho_{\rm scaling}(\vrr)=\frac{1}{\a^3}\rho(\vrr/\a)
\eea
with a scaling factor
\bea
\a=\sqrt{ \frac{\langle \vrr^2 \rangle_{\rm scaling}}{\langle \vrr^2 \rangle}} .
\eea
The actual procedure to determine $\a$ (of p and n) for each case is: 
Firstly we scale the proton density so as to be 
$R_{\rm p}({\rm scaling})=R_{\rm p}({\rm exp})$ 
although is a tiny adjustment, 
secondly we scale the neutron density so as that the $\sigma_{\rm R}$ reproduces 
the data in average with respect to $E_{\rm in}$. 

\section{Results and Discussion}
\label{Results}

\subsection{$^{208}$Pb}
\label{Pb}

\begin{figure}[H]
\begin{center}
 \includegraphics[width=0.4\textwidth,clip]{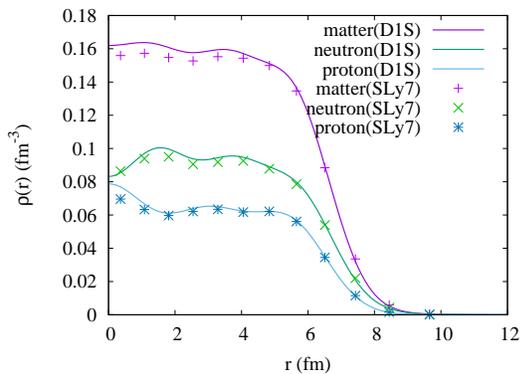}
 \caption{ 
 $r$ dependence of densities,  $\rho_{\rm p}(r)$, $\rho_{\rm n}(r)$, $\rho_{\rm m}(r)$, for $^{208}$Pb calculated with 
 D1S-GHFB+AMP and SLy7-HFB. 
 Dashed curves from the bottom to the top denote the $\rho_{\rm p}(r)$, $\rho_{\rm n}(r)$, $\rho_{\rm m}(r)$ 
 of D1S-GHFB, respectively.  Symbols correspond to the SLy7-HFB densities. 
   }
 \label{Fig-Densities+Pb}
\end{center}
\end{figure}

As a preparation, first we compare two sets of calculated densities in 
Fig.~\ref{Fig-Densities+Pb}. 
The two sets practically coincide in the sense that the effect of the slight 
difference in the deep inside on $\sigma_{\rm R}$ is negligible and the differences 
in the calculated radii (column 1 and 2 in Table~\ref{table1}) are less than 1~$\%$. 

%%%%%%%%%%%%%%%%%%%%%%%
%%%  Figure
%%%%%%%%%%%%%%%%%%%%%%%

%%%%%%%%%%%%%%%%%%%%%%%
%%%  Table
%%%%%%%%%%%%%%%%%%%%%%%
\begin{table}[htbp]
 \caption {Various radii of $^{208}$Pb, given in fm. Column 1 and 2 are the results of 
direct calculations with the Gogny-HFB+AMP and the Skyrme-HFB, respectively. 
Column 3 is taken from Ref.~\cite{Adhikari:2021phr}. Column 4 and 5 are deduced from the renormalized 
densities for p scattering~\cite{Tagami:2020bee} and $^{4}$He scattering (present work), respectively. 
$R_{\rm p}=$ 5.444 fm is taken from Ref.~\cite{PRC.90.067304}.}
 \label{table1}
\begin{center}
\begin{tabular}{lccccc} \hline\hline
        & D1S & SLy7 & PREX II & p & $^{4}$He \\ \hline
$R_{\rm n}$ & 5.580 & 5.619 &      & 5.722$\pm$0.035 & 5.860$\pm$0.146 \\
$R_{\rm p}$ & 5.443 & 5.469 &      & 5.444 & 5.444 \\
$R_{\rm skin}$ & 0.137 & 0.150 & 0.283$\pm$0.071 & 0.278$\pm$0.035 & 0.416$\pm$0.146 \\
$R_{\rm m}$ & 5.526 & 5.560 &      & 5.614$\pm$0.022 & 5.700$\pm$0.146 \\
\hline\hline
\end{tabular}
\end{center}
\end{table}

\begin{figure}[H]
\begin{center}
 \includegraphics[width=0.4\textwidth,clip]{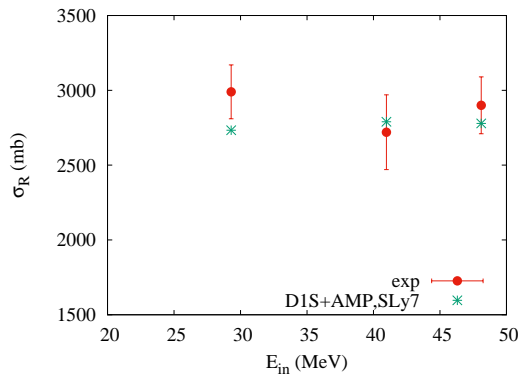}
 \caption{ 
 $E_{\rm in}$ dependence of reaction cross sections $\sigma_{\rm R}$ 
 for $^{4}$He + $^{208}$Pb scattering. 
 Note that $E_{\rm in}$ is the incident energy per nucleon. 
 Asterisks stand for the results of  D1S-GHFB+AMP and SLy7-HFB. 
 The data are taken from Ref.~\cite{Ingemarsson:2000vfz}. 
   }
 \label{Fig-RXsec-p+Pb-1}
\end{center}
\end{figure}

We present the calculated $\sigma_{\rm R}$ in Fig.~\ref{Fig-RXsec-p+Pb-1} 
comparing with the data. 
Adopting the Kyushu chiral $g$-matrix folding model, of which reliability in the energy range 
29.3 $\le E_{\rm in}\le$ 85 MeV has been confirmed~\cite{Tagami:2019svt}, calculated 
$\sigma_{\rm R}$ are 96.6~$\%$ of the data in average. 
This is very similar to the p + $^{208}$Pb result in Ref.~\cite{Tagami:2020bee}, 97$\%$ in 
30 $\le E_{\rm in}\le$ 100 MeV. 

%%%%%%%%%%%%%%%%%%%%%%%
%%%  Figure
%%%%%%%%%%%%%%%%%%%%%%%

According to this observation, we apply the same renormalization procedure, that is, 
we scale the D1S-GHFB+AMP densities so as to reproduce $\sigma_{\rm R}$ for each $E_{\rm in}$ 
under the condition that the proton radius given by the scaled density agrees with the data 
from electron scattering, and take the weighted mean and its error for the resulting $R_{\rm m}$. 
From the resulting $R_{\rm m}=5.700 \pm 0.146$~fm and $R_{\rm p}=5.444$~{\rm fm}, 
we obtain $R_{\rm n}=5.860 \pm 0.146$~fm. 
This leads to $R_{\rm skin}=0.416 \pm 0.146$~fm that is consistent with the PREX II result 
as shown in Table~\ref{table1}. 
The present result for $^{4}$He + $^{208}$Pb, in addition to that for p + $^{208}$Pb 
in Ref.~\cite{Tagami:2020bee}, strongly suggests that the less-determined mean-field parameters in 
the neutron sector tend to lead to smaller $R_{\rm n}$ and consequently thin skins 
at least in heavy nuclei such as $^{208}$Pb. 

This looks consistent with the result of the dispersive optical model analysis, in which 
the single-particle selfenergies are informed by various observed quantities, 
$R_{\rm skin}=0.25 \pm 0.05$~fm~\cite{PRC.101.044303}. 
The remaining unresolved issue is the consistency with the result of the other clean method, 
the electric dipole polarizability $\alpha_{\rm D}$, as explicitly addressed in Ref.~\cite{piekarewicz2021implications}; 
$\alpha_{\rm D}$ obtained from photoabsorption reactions leads to a thin 
$R_{\rm skin}=0.156^{+0.025}_{-0.021}$~fm~\cite{PRL.107.062502} through the correlation~\cite{PhysRevC.81.051303}. 

\subsection{$^{116,120,124}$Sn}
\label{Sn}

\begin{figure}[H]
\begin{center}
 \includegraphics[width=0.4\textwidth,clip]{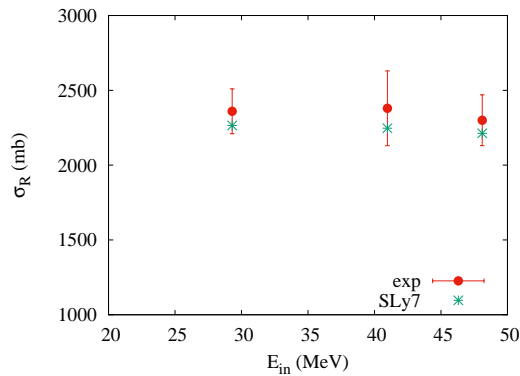}
 \caption{ 
 $E_{\rm in}$ dependence of reaction cross sections $\sigma_{\rm R}$ 
 for $^{4}$He + $^{120}$Sn scattering. 
 Asterisks show the results of  SLy7-HFB. 
 The data are taken from Ref.~\cite{Ingemarsson:2000vfz}. 
   }
 \label{Fig-RXsec-p+Sn120}
\end{center}
\end{figure}

In order to see to what extent the picture presented above holds in lighter nuclides, 
we study stable Sn isotopes in this subsection. 
Figure~\ref{Fig-RXsec-p+Sn120} presents the experimental and calculated $\sigma_{\rm R}$ 
for $^{120}$Sn. 
The latter is smaller than the central value of the former although located within the error bar at all three $E_{\rm in}$. 
The direct calculation with the SLy7 parameter set gives $R_{\rm skin}=$ 0.123 fm as 
in column 1 in Table~\ref{table2}. 
This is slightly smaller than the HF+BCS result with the SLy4 set reported in Ref.~\cite{PhysRevC.76.044322}. 
Column 2 reports the result of the same renormalization procedure as above, adopting 
$R_{\rm p}=$ 4.583 fm~\cite{ADNDT.99.69}. 
The resulting skin is too thick. 

%%%%%%%%%%%%%%%%%%%%%%%
%%%  Figure
%%%%%%%%%%%%%%%%%%%%%%%

%%%%%%%%%%%%%%%%%%%%%%%
%%%  Table
%%%%%%%%%%%%%%%%%%%%%%%
\begin{table}[htbp]
 \caption {Various radii of $^{120}$Sn, given in fm. Column 1 is the result of 
the direct calculation with the Skyrme-HFB. 
Column 2 is that of the same renormalization procedure applied to the 
data~\cite{Ingemarsson:2000vfz} with $R_{\rm p}=$ 4.583 fm, determined from the charge density data~\cite{ADNDT.99.69}. 
Column 3 and 4 are taken from Refs.~\cite{Krasznahorkay:1999zz,Hashimoto:2015ema}.}
 \label{table2}
\begin{center}
\begin{tabular}{lcccc} \hline\hline
            & SLy7 & $^{4}$He & Krasznahorkay & Hashimoto \\ \hline
$R_{\rm n}$ & 4.719 & 4.959$\pm$0.140 &     &      \\
$R_{\rm p}$ & 4.595 & 4.583           &     &      \\
$R_{\rm skin}$ & 0.123 & 0.377$\pm$0.140 & 0.18$\pm$0.07 & 0.148$\pm$0.034 \\
$R_{\rm m}$ & 4.668 & 4.806$\pm$0.140 &     &      \\
\hline\hline
\end{tabular}
\end{center}
\end{table}

Then we consult other experimental information obtained by dipole resonances. 
The first one is given by the spin-dipole resonance excited by the ($^{3}$He,t) 
reaction~\cite{Krasznahorkay:1999zz}. 
A model-dependent value $R_{\rm skin}= 0.18\pm0.07$ fm is given by normalizing 
to a theoretical result. 
The second one is given through the correlation with $\alpha_{\rm D}$~\cite{PhysRevC.81.051303} 
obtained by the ($\vec{p}$,$\vec{p}\,'$) reaction~\cite{Hashimoto:2015ema}. 
The authors conclude $R_{\rm skin}= 0.148\pm0.034$ fm. 
These are summarized in Table~\ref{table2}. 
Although the correlation between $R_{\rm skin}$ and $\alpha_{\rm D}$ is 
argued not to be consistent with relativistic mean-field calculations in 
$^{208}$Pb~\cite{piekarewicz2021implications}, in the present case of $^{120}$Sn it is consistent at least 
with the selected Skyrme parameter sets. On the other hand, the performance 
of the present prescription applied to the $\sigma_{\rm R}$ data is not good. 
Therefore we suspect that the $\sigma_{\rm R}$ data contain some error. 

\begin{figure}[H]
\begin{center}
 \includegraphics[width=0.4\textwidth,clip]{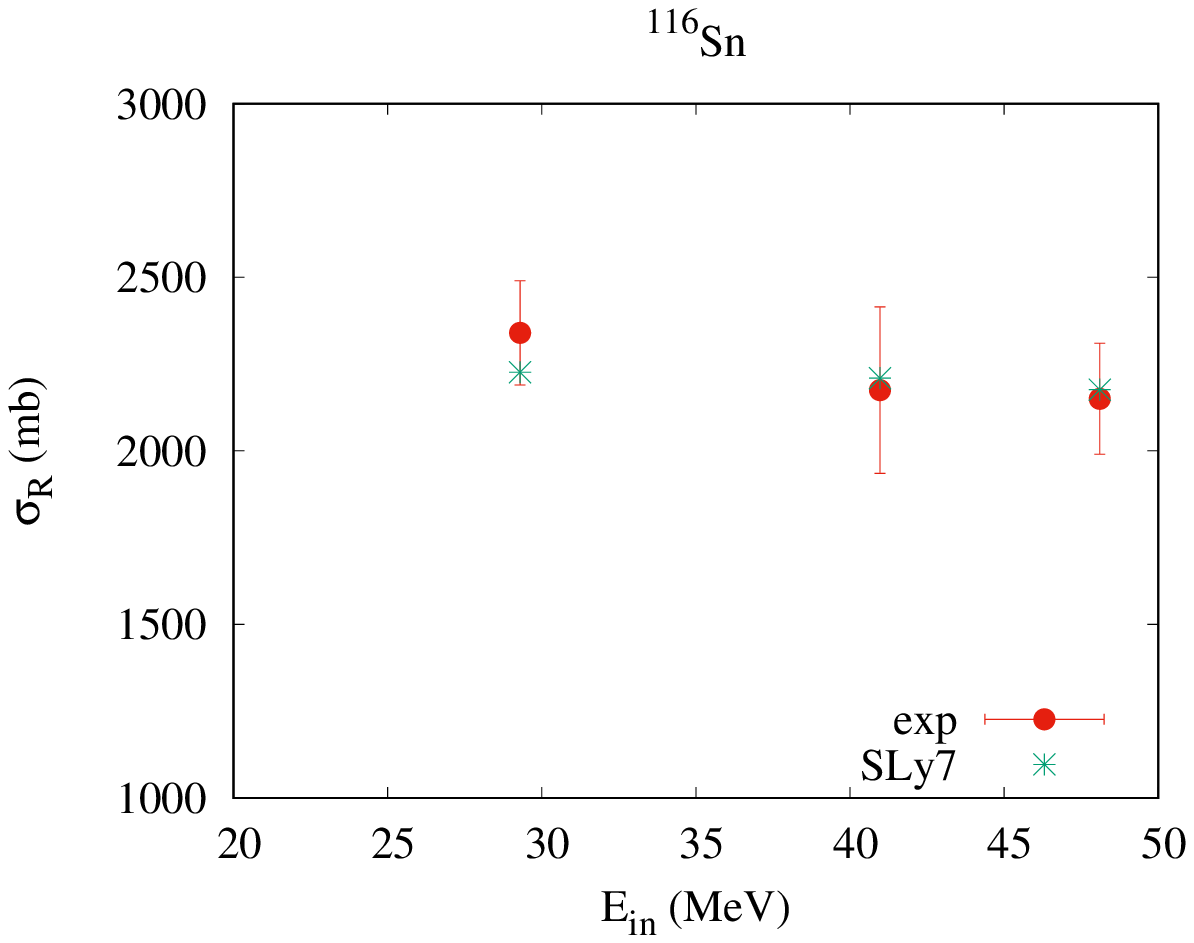}
 \includegraphics[width=0.4\textwidth,clip]{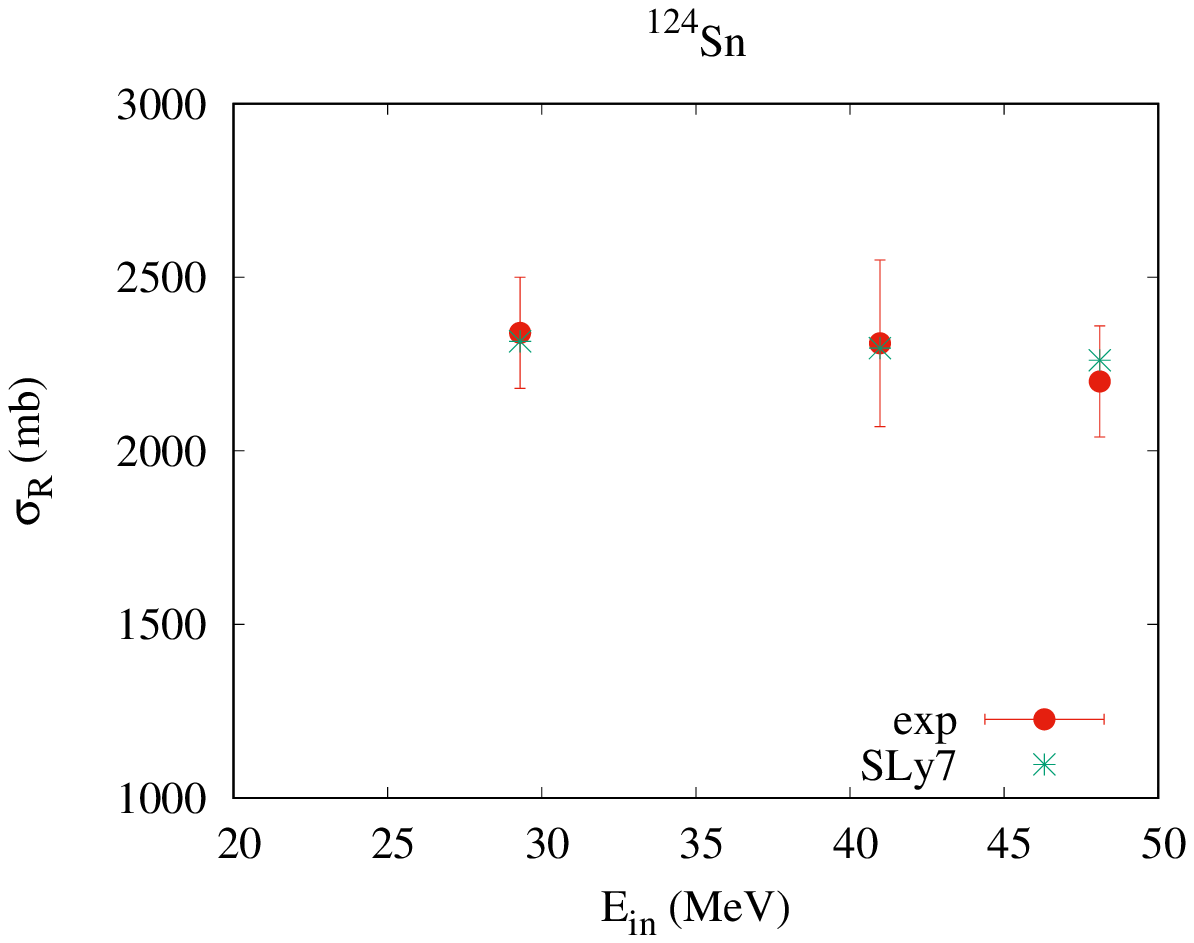}
 \includegraphics[width=0.4\textwidth,clip]{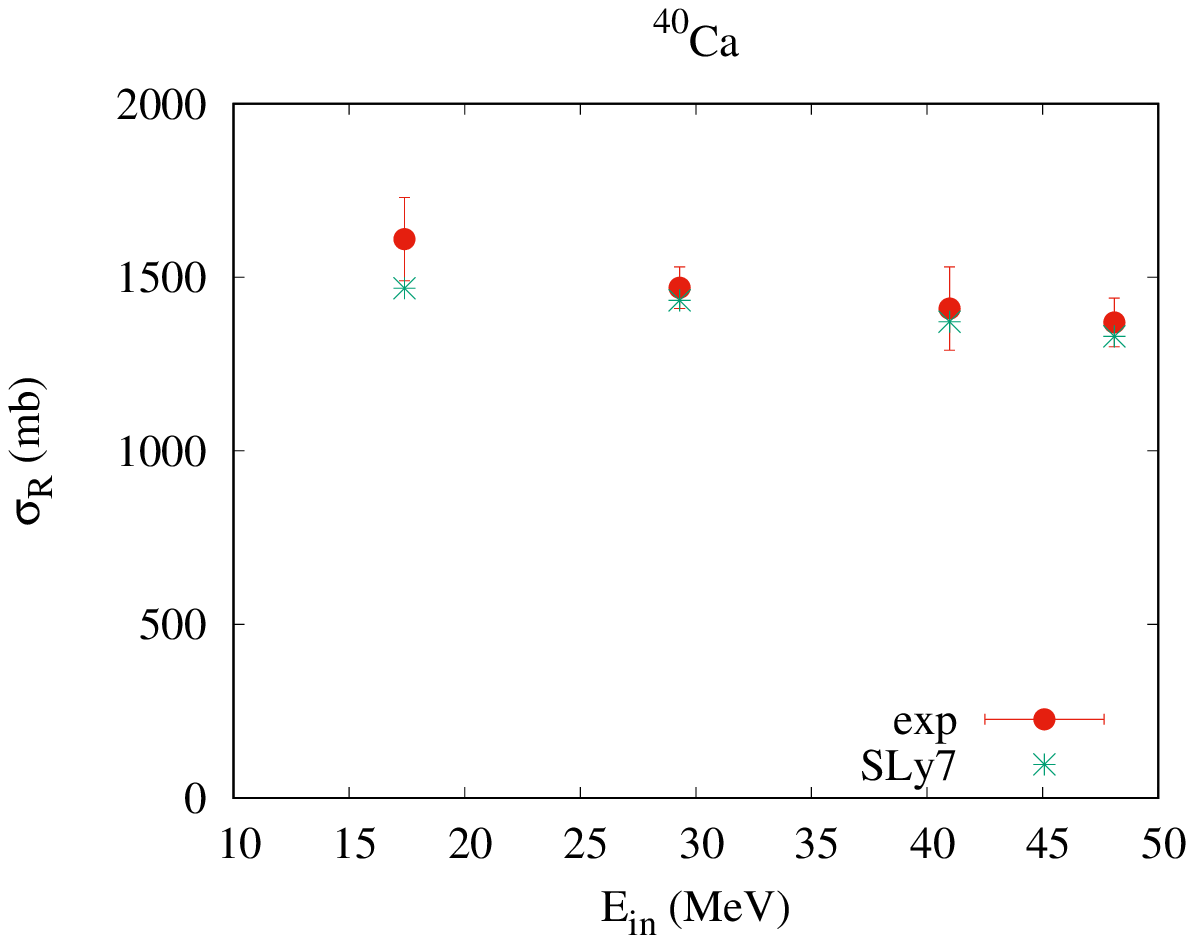}
 \caption{Three panels from the top to the bottom show 
 $E_{\rm in}$ dependence of reaction cross sections $\sigma_{\rm R}$ 
 for $^{4}$He + $^{116,124}$Sn, and $^{40}$Ca scattering, respectively. 
 Asterisks show the results of  SLy7-HFB. 
 The data are taken from Ref.~\cite{Ingemarsson:2000vfz}. 
   }
 \label{Fig-RXsec-p+Sn116-124Ca40}
\end{center}
\end{figure}

Next we examine $^{116,124}$Sn. 
Many data extracted from various methods, 
which are presented in Fig.4 of Ref.~\cite{PhysRevC.76.044322}, are available in addition to the 
$\sigma_{\rm R}$ data of present interest. 
Our results are shown in Fig.~\ref{Fig-RXsec-p+Sn116-124Ca40} and Table~\ref{table3}. 
That of $^{124}$Sn looks consistent with other experimental and theoretical results, 
but in the $^{116}$Sn case $R_{\rm skin}$ extracted from $\sigma_{\rm R}$ is 
evidently too large as in the $^{120}$Sn case above. 

%%%%%%%%%%%%%%%%%%%%%%%
%%%  Figure
%%%%%%%%%%%%%%%%%%%%%%%

%%%%%%%%%%%%%%%%%%%%%%%
%%%  Table
%%%%%%%%%%%%%%%%%%%%%%%
\begin{table*}[htbp]
 \caption {Various radii of $^{116,124}$Sn, and $^{40}$Ca, given in fm. Column 1, 4, and 7 are the results of 
the direct calculations with the Skyrme-HFB. 
Column 2, 5, and 8 are those of the same renormalization procedure applied to the 
data~\cite{Ingemarsson:2000vfz} with $R_{\rm p}=$ 4.554, 4.606, and 3.378 fm, determined from the charge density data~\cite{ADNDT.99.69}. 
Column 3, 6, and 9 are taken from Refs.~\cite{Krasznahorkay:1999zz,zenihiro2018direct}.}
 \label{table3}
\begin{center}
\begin{tabular}{llcclccrcr} \hline\hline
 & $^{116}$Sn & & & $^{124}$Sn & & & $^{40}$Ca & & \\
 & SLy7 & $^{4}$He & Krasznahorkay & SLy7 & $^{4}$He & Krasznahorkay & SLy7 & $^{4}$He & Zenihiro \\ \hline
$R_{\rm n}$    & 4.654 & 4.796$\pm$0.140 &               & 4.779 & 4.785$\pm$0.142 &               & 3.359  & 3.343$\pm$0.075  & 3.375$^{+0.022}_{-0.023}$  \\
$R_{\rm p}$    & 4.565 & 4.554           &               & 4.623 & 4.606           &               & 3.406  & 3.378            & 3.385         \\
$R_{\rm skin}$ & 0.089 & 0.242$\pm$0.140 & 0.12$\pm$0.06 & 0.155 & 0.180$\pm$0.142 & 0.19$\pm$0.07 & -0.043 & -0.035$\pm$0.075 & -0.010$^{+0.022}_{-0.024}$ \\
$R_{\rm m}$    & 4.616 & 4.693$\pm$0.140 &               & 4.717 & 4.714$\pm$0.142 &               & 3.381  & 3.361$\pm$0.075  &               \\
\hline\hline
\end{tabular}
\end{center}
\end{table*}

\subsection{$^{40}$Ca}
\label{Ca}

As the last example, we take $^{40}$Ca in order to see whether the present method 
is applicable also to the case with $R_{\rm skin}\lsim$ 0. 
$^{48}$Ca will be studied separately elsewhere. 
The results are shown in Fig.~\ref{Fig-RXsec-p+Sn116-124Ca40} and Table~\ref{table3}. 
In Table~\ref{table3}, our result is compared with that deduced from the recent 
proton elastic scattering~\cite{zenihiro2018direct}. 
These indicate that the present method works well also for the $R_{\rm skin}\lsim$ 0 
case, at the same time, indicate that the mean-field parameters of the neutron 
sector is more reliable than in heavier cases. 

\section{Summary}
\label{Summary}

\begin{figure}[H]
\begin{center}
  \includegraphics[width=0.4\textwidth,clip]{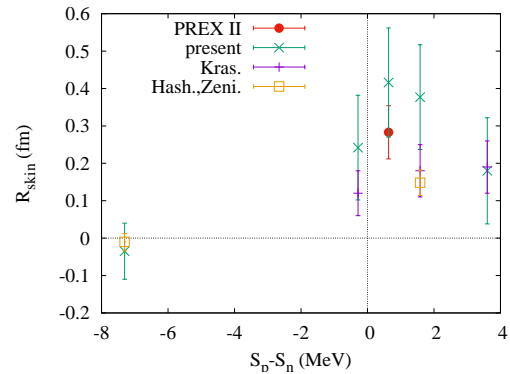}
 \caption{ 
 Skin thicknesses deduced in the present and other works are summarized as a function of 
 the difference between the proton and neutron separation energies: 
 -7.31, -0.28, 0.64, 1.58, and 3.60 MeV for 
 $^{40}$Ca, $^{116}$Sn, $^{208}$Pb, $^{120}$Sn, and $^{124}$Sn, respectively. 
 The data are taken from 
 Refs.~\cite{Adhikari:2021phr,Krasznahorkay:1999zz,Hashimoto:2015ema,zenihiro2018direct}. 
   }
 \label{Fig-skins}
\end{center}
\end{figure}

Based on the studies of reaction cross sections that confirm the double 
folding model with the Kyushu chiral $g$-matrix at each $E_{\rm in}$ 
and the Gogny and Skyrme HFB, we examined to deduce the neutron skin thicknesses 
of $^{208}$Pb, $^{116,120,124}$Sn, and $^{40}$Ca. 
First we found that the present model gives 3.4~$\%$ smaller cross sections for 
$^{4}$He + $^{208}$Pb, similarly to 3 $\%$ in the single folding case of 
p + $^{208}$Pb in a preceding work. 
We attributed the origin of these deviations to less-confirmed mean-field 
parameters for neutrons in heavy nuclei, and renormalized the HFB densities. 
Then the nuclear matter radii deduced from cross sections lead to skin thicknesses 
by confronting precision proton radii. 
The result is consistent with that of PREX II. 
Then we applied the method to lighter nuclides. 
Among stable Sn isotopes, for which the $\sigma_{\rm R}$ data show rather large 
error, this method leads to thicker skins in $^{116,120}$Sn. 
This indicates that other observables should also be examined. 
For $^{40}$Ca in which mean-field parameters are thought to be relatively well 
determined and $R_{\rm skin}\lsim$ 0, this method works well. 
We summarize our numerical results for the five nuclides in Fig.~\ref{Fig-skins} as a function 
of the separation energy difference.

Using the fitted relation between the skin thickness of $^{208}$Pb and the slope 
parameter of symmetry energy, 
$R_{\rm skin}^{208}=0.101+0.00147L$~\cite{Roca-Maza:2011qcr}, 
our result $R_{\rm skin}^{208}=$ 0.416$\pm $0.146~fm and 
$R_{\rm skin}^{208}({\rm PREX~II}) = 0.283\pm 0.071\,{\rm fm}$ 
lead to $L=115 - 313.6$~MeV and $L=75.5 - 172.1$~MeV, respectively. 
These values support stiffer EoSs and exclude APR ($L\approx 40$~MeV). 
This is the point we found out through the present study in relation to the symmetry energy.

%%%%%%%%%%%%%%%%%%%%%%%
%%%  Figure
%%%%%%%%%%%%%%%%%%%%%%%

%%%%%%%%%%%%%%%%%%%%%%%%%%%%%%%%%%%%%%%%%%%%%%%%%%%%%%%%%%%%%%%%%%%%%%%%%%%%%%%%
%%%%% Acknowledgments 
%%%%%%%%%%%%%%%%%%%%%%%%%%%%%%%%%%%%%%%%%%%%%%%%%%%%%%%%%%%%%%%%%%%%%%%%%%%%%%%%
\noindent
\begin{acknowledgments}
We would like to thank Dr. Toyokawa for providing his code. 
\end{acknowledgments}

%%%%%%%%%%%%%%%%%%%%%%%%%%%%%%%%%%%%%%%%%%%%%%%%%%%%%%%%%%%%%%%%%%%%%%%%%%%%%%%%%%%%% References 
%%%%%%%%%%%%%%%%%%%%%%%%%%%%%%%%%%%%%%%%%%%%%%%%%%%%%%%%%%%%%%%%%%%%%%%%%%%%%%%%

%%%%%%%%%%%%%%%%%%%%%%%%%%%%%%%%%%%%%%%%%%%%%%%%%%%%%%%%%%%%%%%%%%%%%%%%%%%%%%%%%%%%% References 
%%%%%%%%%%%%%%%%%%%%%%%%%%%%%%%%%%%%%%%%%%%%%%%%%%%%%%%%%%%%%%%%%%%%%%%%%%%%%%%%

%%%%%%%%%%%%%%%%%%%%%%%%%%%%%%%%%%%%%%%%%%

% Create the reference section using BibTeX:
\bibliography{Folding-v5_MM}

%merlin.mbs apsrev4-1.bst 2010-07-25 4.21a (PWD, AO, DPC) hacked
%Control: key (0)
%Control: author (8) initials jnrlst
%Control: editor formatted (1) identically to author
%Control: production of article title (-1) disabled
%Control: page (0) single
%Control: year (1) truncated
%Control: production of eprint (0) enabled
\begin{thebibliography}{26}%
\makeatletter
\providecommand \@ifxundefined [1]{%
 \@ifx{#1\undefined}
}%
\providecommand \@ifnum [1]{%
 \ifnum #1\expandafter \@firstoftwo
 \else \expandafter \@secondoftwo
 \fi
}%
\providecommand \@ifx [1]{%
 \ifx #1\expandafter \@firstoftwo
 \else \expandafter \@secondoftwo
 \fi
}%
\providecommand \natexlab [1]{#1}%
\providecommand \enquote  [1]{``#1''}%
\providecommand \bibnamefont  [1]{#1}%
\providecommand \bibfnamefont [1]{#1}%
\providecommand \citenamefont [1]{#1}%
\providecommand \href@noop [0]{\@secondoftwo}%
\providecommand \href [0]{\begingroup \@sanitize@url \@href}%
\providecommand \@href[1]{\@@startlink{#1}\@@href}%
\providecommand \@@href[1]{\endgroup#1\@@endlink}%
\providecommand \@sanitize@url [0]{\catcode `\\12\catcode `\$12\catcode
  `\&12\catcode `\#12\catcode `\^12\catcode `\_12\catcode `\%12\relax}%
\providecommand \@@startlink[1]{}%
\providecommand \@@endlink[0]{}%
\providecommand \url  [0]{\begingroup\@sanitize@url \@url }%
\providecommand \@url [1]{\endgroup\@href {#1}{\urlprefix }}%
\providecommand \urlprefix  [0]{URL }%
\providecommand \Eprint [0]{\href }%
\providecommand \doibase [0]{http://dx.doi.org/}%
\providecommand \selectlanguage [0]{\@gobble}%
\providecommand \bibinfo  [0]{\@secondoftwo}%
\providecommand \bibfield  [0]{\@secondoftwo}%
\providecommand \translation [1]{[#1]}%
\providecommand \BibitemOpen [0]{}%
\providecommand \bibitemStop [0]{}%
\providecommand \bibitemNoStop [0]{.\EOS\space}%
\providecommand \EOS [0]{\spacefactor3000\relax}%
\providecommand \BibitemShut  [1]{\csname bibitem#1\endcsname}%
\let\auto@bib@innerbib\@empty
%</preamble>
\bibitem [{\citenamefont {Horowitz}\ \emph {et~al.}(2001)\citenamefont
  {Horowitz}, \citenamefont {Pollock}, \citenamefont {Souder},\ and\
  \citenamefont {Michaels}}]{PRC.63.025501}%
  \BibitemOpen
  \bibfield  {author} {\bibinfo {author} {\bibfnamefont {C.~J.}\ \bibnamefont
  {Horowitz}}, \bibinfo {author} {\bibfnamefont {S.~J.}\ \bibnamefont
  {Pollock}}, \bibinfo {author} {\bibfnamefont {P.~A.}\ \bibnamefont {Souder}},
  \ and\ \bibinfo {author} {\bibfnamefont {R.}~\bibnamefont {Michaels}},\
  }\href {\doibase 10.1103/PhysRevC.63.025501} {\bibfield  {journal} {\bibinfo
  {journal} {Phys. Rev. C}\ }\textbf {\bibinfo {volume} {63}},\ \bibinfo
  {pages} {025501} (\bibinfo {year} {2001})}\BibitemShut {NoStop}%
\bibitem [{\citenamefont {Adhikari}\ \emph {et~al.}(2021)\citenamefont
  {Adhikari} \emph {et~al.}}]{Adhikari:2021phr}%
  \BibitemOpen
  \bibfield  {author} {\bibinfo {author} {\bibfnamefont {D.}~\bibnamefont
  {Adhikari}} \emph {et~al.} (\bibinfo {collaboration} {PREX}),\ }\href
  {\doibase 10.1103/PhysRevLett.126.172502} {\bibfield  {journal} {\bibinfo
  {journal} {Phys. Rev. Lett.}\ }\textbf {\bibinfo {volume} {126}},\ \bibinfo
  {pages} {172502} (\bibinfo {year} {2021})},\ \Eprint
  {http://arxiv.org/abs/2102.10767} {arXiv:2102.10767 [nucl-ex]} \BibitemShut
  {NoStop}%
\bibitem [{\citenamefont {Akmal}\ \emph {et~al.}(1998)\citenamefont {Akmal},
  \citenamefont {Pandharipande},\ and\ \citenamefont
  {Ravenhall}}]{Akmal:1998cf}%
  \BibitemOpen
  \bibfield  {author} {\bibinfo {author} {\bibfnamefont {A.}~\bibnamefont
  {Akmal}}, \bibinfo {author} {\bibfnamefont {V.~R.}\ \bibnamefont
  {Pandharipande}}, \ and\ \bibinfo {author} {\bibfnamefont {D.~G.}\
  \bibnamefont {Ravenhall}},\ }\href {\doibase 10.1103/PhysRevC.58.1804}
  {\bibfield  {journal} {\bibinfo  {journal} {Phys. Rev. C}\ }\textbf {\bibinfo
  {volume} {58}},\ \bibinfo {pages} {1804} (\bibinfo {year} {1998})},\ \Eprint
  {http://arxiv.org/abs/nucl-th/9804027} {arXiv:nucl-th/9804027} \BibitemShut
  {NoStop}%
\bibitem [{\citenamefont {Li}\ \emph {et~al.}(2014)\citenamefont {Li},
  \citenamefont {Ramos}, \citenamefont {Verde},\ and\ \citenamefont
  {Vidana}}]{Li:2014oda}%
  \BibitemOpen
  \bibfield  {author} {\bibinfo {author} {\bibfnamefont {B.-A.}\ \bibnamefont
  {Li}}, \bibinfo {author} {\bibfnamefont {A.}~\bibnamefont {Ramos}}, \bibinfo
  {author} {\bibfnamefont {G.}~\bibnamefont {Verde}}, \ and\ \bibinfo {author}
  {\bibfnamefont {I.}~\bibnamefont {Vidana}},\ }\href {\doibase
  10.1140/epja/i2014-14009-x} {\bibfield  {journal} {\bibinfo  {journal} {Eur.
  Phys. J. A}\ }\textbf {\bibinfo {volume} {50}},\ \bibinfo {pages} {9}
  (\bibinfo {year} {2014})}\BibitemShut {NoStop}%
\bibitem [{\citenamefont {Toyokawa}\ \emph {et~al.}(2018)\citenamefont
  {Toyokawa}, \citenamefont {Yahiro}, \citenamefont {Matsumoto},\ and\
  \citenamefont {Kohno}}]{Toyokawa:2017pdd}%
  \BibitemOpen
  \bibfield  {author} {\bibinfo {author} {\bibfnamefont {M.}~\bibnamefont
  {Toyokawa}}, \bibinfo {author} {\bibfnamefont {M.}~\bibnamefont {Yahiro}},
  \bibinfo {author} {\bibfnamefont {T.}~\bibnamefont {Matsumoto}}, \ and\
  \bibinfo {author} {\bibfnamefont {M.}~\bibnamefont {Kohno}},\ }\href
  {\doibase 10.1093/ptep/pty001} {\bibfield  {journal} {\bibinfo  {journal}
  {PTEP}\ }\textbf {\bibinfo {volume} {2018}},\ \bibinfo {pages} {023D03}
  (\bibinfo {year} {2018})},\ \Eprint {http://arxiv.org/abs/1712.07033}
  {arXiv:1712.07033 [nucl-th]} \BibitemShut {NoStop}%
\bibitem [{\citenamefont {Kohno}(2013)}]{PRC.88.064005}%
  \BibitemOpen
  \bibfield  {author} {\bibinfo {author} {\bibfnamefont {M.}~\bibnamefont
  {Kohno}},\ }\href {\doibase 10.1103/PhysRevC.88.064005} {\bibfield  {journal}
  {\bibinfo  {journal} {Phys. Rev. C}\ }\textbf {\bibinfo {volume} {88}},\
  \bibinfo {pages} {064005} (\bibinfo {year} {2013})}\BibitemShut {NoStop}%
\bibitem [{\citenamefont
  {https://sites.google.com/view/kyushu-nucl-th/misc/parameter-sets-of-kyushu-chiral-g
  matrix}()}]{localized}%
  \BibitemOpen
  \bibfield  {author} {\bibinfo {author} {\bibnamefont
  {https://sites.google.com/view/kyushu-nucl-th/misc/parameter-sets-of-kyushu-chiral-g
  matrix}},\ }\href@noop {} {}\BibitemShut {NoStop}%
\bibitem [{\citenamefont {de~Vries}\ \emph {et~al.}(1987)\citenamefont
  {de~Vries}, \citenamefont {de~Jager},\ and\ \citenamefont
  {de~Vries}}]{C12-density}%
  \BibitemOpen
  \bibfield  {author} {\bibinfo {author} {\bibfnamefont {H.}~\bibnamefont
  {de~Vries}}, \bibinfo {author} {\bibfnamefont {C.~W.}\ \bibnamefont
  {de~Jager}}, \ and\ \bibinfo {author} {\bibfnamefont {C.}~\bibnamefont
  {de~Vries}},\ }\href@noop {} {\bibfield  {journal} {\bibinfo  {journal} {At.
  Data Nucl. Data Tables}\ }\textbf {\bibinfo {volume} {36}},\ \bibinfo {pages}
  {495} (\bibinfo {year} {1987})}\BibitemShut {NoStop}%
\bibitem [{\citenamefont {Tagami}\ \emph {et~al.}(2020)\citenamefont {Tagami},
  \citenamefont {Tanaka}, \citenamefont {Takechi}, \citenamefont {Fukuda},\
  and\ \citenamefont {Yahiro}}]{Tagami:2019svt}%
  \BibitemOpen
  \bibfield  {author} {\bibinfo {author} {\bibfnamefont {S.}~\bibnamefont
  {Tagami}}, \bibinfo {author} {\bibfnamefont {M.}~\bibnamefont {Tanaka}},
  \bibinfo {author} {\bibfnamefont {M.}~\bibnamefont {Takechi}}, \bibinfo
  {author} {\bibfnamefont {M.}~\bibnamefont {Fukuda}}, \ and\ \bibinfo {author}
  {\bibfnamefont {M.}~\bibnamefont {Yahiro}},\ }\href {\doibase
  10.1103/PhysRevC.101.014620} {\bibfield  {journal} {\bibinfo  {journal}
  {Phys. Rev. C}\ }\textbf {\bibinfo {volume} {101}},\ \bibinfo {pages}
  {014620} (\bibinfo {year} {2020})},\ \Eprint
  {http://arxiv.org/abs/1911.05417} {arXiv:1911.05417 [nucl-th]} \BibitemShut
  {NoStop}%
\bibitem [{\citenamefont {Tagami}\ \emph {et~al.}(2021)\citenamefont {Tagami},
  \citenamefont {Wakasa}, \citenamefont {Matsui}, \citenamefont {Yahiro},\ and\
  \citenamefont {Takechi}}]{Tagami:2020bee}%
  \BibitemOpen
  \bibfield  {author} {\bibinfo {author} {\bibfnamefont {S.}~\bibnamefont
  {Tagami}}, \bibinfo {author} {\bibfnamefont {T.}~\bibnamefont {Wakasa}},
  \bibinfo {author} {\bibfnamefont {J.}~\bibnamefont {Matsui}}, \bibinfo
  {author} {\bibfnamefont {M.}~\bibnamefont {Yahiro}}, \ and\ \bibinfo {author}
  {\bibfnamefont {M.}~\bibnamefont {Takechi}},\ }\href {\doibase
  10.1103/PhysRevC.104.024606} {\bibfield  {journal} {\bibinfo  {journal}
  {Phys. Rev. C}\ }\textbf {\bibinfo {volume} {104}},\ \bibinfo {pages}
  {024606} (\bibinfo {year} {2021})},\ \Eprint
  {http://arxiv.org/abs/2010.02450} {arXiv:2010.02450 [nucl-th]} \BibitemShut
  {NoStop}%
\bibitem [{\citenamefont {Schunck}\ \emph {et~al.}(2017)\citenamefont {Schunck}
  \emph {et~al.}}]{Schunck:2016uvm}%
  \BibitemOpen
  \bibfield  {author} {\bibinfo {author} {\bibfnamefont {N.}~\bibnamefont
  {Schunck}} \emph {et~al.},\ }\href {\doibase 10.1016/j.cpc.2017.03.007}
  {\bibfield  {journal} {\bibinfo  {journal} {Comput. Phys. Commun.}\ }\textbf
  {\bibinfo {volume} {216}},\ \bibinfo {pages} {145} (\bibinfo {year}
  {2017})},\ \Eprint {http://arxiv.org/abs/1612.05314} {arXiv:1612.05314
  [nucl-th]} \BibitemShut {NoStop}%
\bibitem [{\citenamefont {Chabanat}\ \emph {et~al.}(1998)\citenamefont
  {Chabanat}, \citenamefont {Bonche}, \citenamefont {Haensel}, \citenamefont
  {Meyer},\ and\ \citenamefont {Schaeffer}}]{Chabanat:1997un}%
  \BibitemOpen
  \bibfield  {author} {\bibinfo {author} {\bibfnamefont {E.}~\bibnamefont
  {Chabanat}}, \bibinfo {author} {\bibfnamefont {P.}~\bibnamefont {Bonche}},
  \bibinfo {author} {\bibfnamefont {P.}~\bibnamefont {Haensel}}, \bibinfo
  {author} {\bibfnamefont {J.}~\bibnamefont {Meyer}}, \ and\ \bibinfo {author}
  {\bibfnamefont {R.}~\bibnamefont {Schaeffer}},\ }\href {\doibase
  10.1016/S0375-9474(98)00180-8} {\bibfield  {journal} {\bibinfo  {journal}
  {Nucl. Phys. A}\ }\textbf {\bibinfo {volume} {635}},\ \bibinfo {pages} {231}
  (\bibinfo {year} {1998})},\ \bibinfo {note} {[Erratum: Nucl.Phys.A 643,
  441--441 (1998)]}\BibitemShut {NoStop}%
\bibitem [{\citenamefont {Minomo}\ \emph {et~al.}(2010)\citenamefont {Minomo},
  \citenamefont {Ogata}, \citenamefont {Kohno}, \citenamefont {Shimizu},\ and\
  \citenamefont {Yahiro}}]{Minomo:2009ds}%
  \BibitemOpen
  \bibfield  {author} {\bibinfo {author} {\bibfnamefont {K.}~\bibnamefont
  {Minomo}}, \bibinfo {author} {\bibfnamefont {K.}~\bibnamefont {Ogata}},
  \bibinfo {author} {\bibfnamefont {M.}~\bibnamefont {Kohno}}, \bibinfo
  {author} {\bibfnamefont {Y.~R.}\ \bibnamefont {Shimizu}}, \ and\ \bibinfo
  {author} {\bibfnamefont {M.}~\bibnamefont {Yahiro}},\ }\href {\doibase
  10.1088/0954-3899/37/8/085011} {\bibfield  {journal} {\bibinfo  {journal} {J.
  Phys. G}\ }\textbf {\bibinfo {volume} {37}},\ \bibinfo {pages} {085011}
  (\bibinfo {year} {2010})},\ \Eprint {http://arxiv.org/abs/0911.1184}
  {arXiv:0911.1184 [nucl-th]} \BibitemShut {NoStop}%
\bibitem [{\citenamefont {Egashira}\ \emph {et~al.}(2014)\citenamefont
  {Egashira}, \citenamefont {Minomo}, \citenamefont {Toyokawa}, \citenamefont
  {Matsumoto},\ and\ \citenamefont {Yahiro}}]{PRC.89.064611}%
  \BibitemOpen
  \bibfield  {author} {\bibinfo {author} {\bibfnamefont {K.}~\bibnamefont
  {Egashira}}, \bibinfo {author} {\bibfnamefont {K.}~\bibnamefont {Minomo}},
  \bibinfo {author} {\bibfnamefont {M.}~\bibnamefont {Toyokawa}}, \bibinfo
  {author} {\bibfnamefont {T.}~\bibnamefont {Matsumoto}}, \ and\ \bibinfo
  {author} {\bibfnamefont {M.}~\bibnamefont {Yahiro}},\ }\href {\doibase
  10.1103/PhysRevC.89.064611} {\bibfield  {journal} {\bibinfo  {journal} {Phys.
  Rev. C}\ }\textbf {\bibinfo {volume} {89}},\ \bibinfo {pages} {064611}
  (\bibinfo {year} {2014})}\BibitemShut {NoStop}%
\bibitem [{\citenamefont {Jones}\ and\ \citenamefont
  {Brown}(2014)}]{PRC.90.067304}%
  \BibitemOpen
  \bibfield  {author} {\bibinfo {author} {\bibfnamefont {A.~B.}\ \bibnamefont
  {Jones}}\ and\ \bibinfo {author} {\bibfnamefont {B.~A.}\ \bibnamefont
  {Brown}},\ }\href {\doibase 10.1103/PhysRevC.90.067304} {\bibfield  {journal}
  {\bibinfo  {journal} {Phys. Rev. C}\ }\textbf {\bibinfo {volume} {90}},\
  \bibinfo {pages} {067304} (\bibinfo {year} {2014})}\BibitemShut {NoStop}%
\bibitem [{\citenamefont {Ingemarsson}\ \emph {et~al.}(2000)\citenamefont
  {Ingemarsson} \emph {et~al.}}]{Ingemarsson:2000vfz}%
  \BibitemOpen
  \bibfield  {author} {\bibinfo {author} {\bibfnamefont {A.}~\bibnamefont
  {Ingemarsson}} \emph {et~al.},\ }\href {\doibase
  10.1016/S0375-9474(00)00200-1} {\bibfield  {journal} {\bibinfo  {journal}
  {Nucl. Phys. A}\ }\textbf {\bibinfo {volume} {676}},\ \bibinfo {pages} {3}
  (\bibinfo {year} {2000})}\BibitemShut {NoStop}%
\bibitem [{\citenamefont {Atkinson}\ \emph {et~al.}(2020)\citenamefont
  {Atkinson}, \citenamefont {Mahzoon}, \citenamefont {Keim}, \citenamefont
  {Bordelon}, \citenamefont {Pruitt}, \citenamefont {Charity},\ and\
  \citenamefont {Dickhoff}}]{PRC.101.044303}%
  \BibitemOpen
  \bibfield  {author} {\bibinfo {author} {\bibfnamefont {M.~C.}\ \bibnamefont
  {Atkinson}}, \bibinfo {author} {\bibfnamefont {M.~H.}\ \bibnamefont
  {Mahzoon}}, \bibinfo {author} {\bibfnamefont {M.~A.}\ \bibnamefont {Keim}},
  \bibinfo {author} {\bibfnamefont {B.~A.}\ \bibnamefont {Bordelon}}, \bibinfo
  {author} {\bibfnamefont {C.~D.}\ \bibnamefont {Pruitt}}, \bibinfo {author}
  {\bibfnamefont {R.~J.}\ \bibnamefont {Charity}}, \ and\ \bibinfo {author}
  {\bibfnamefont {W.~H.}\ \bibnamefont {Dickhoff}},\ }\href {\doibase
  10.1103/PhysRevC.101.044303} {\bibfield  {journal} {\bibinfo  {journal}
  {Phys. Rev. C}\ }\textbf {\bibinfo {volume} {101}},\ \bibinfo {pages}
  {044303} (\bibinfo {year} {2020})}\BibitemShut {NoStop}%
\bibitem [{\citenamefont {Piekarewicz}(2021)}]{piekarewicz2021implications}%
  \BibitemOpen
  \bibfield  {author} {\bibinfo {author} {\bibfnamefont {J.}~\bibnamefont
  {Piekarewicz}},\ }\href@noop {} {\enquote {\bibinfo {title} {Implications of
  prex-2 on the electric dipole polarizability of neutron rich nuclei},}\ }
  (\bibinfo {year} {2021}),\ \Eprint {http://arxiv.org/abs/2105.13452}
  {arXiv:2105.13452 [nucl-th]} \BibitemShut {NoStop}%
\bibitem [{\citenamefont {Tamii}\ \emph {et~al.}(2011)\citenamefont {Tamii},
  \citenamefont {Poltoratska}, \citenamefont {von Neumann-Cosel}, \citenamefont
  {Fujita}, \citenamefont {Adachi}, \citenamefont {Bertulani}, \citenamefont
  {Carter} \emph {et~al.}}]{PRL.107.062502}%
  \BibitemOpen
  \bibfield  {author} {\bibinfo {author} {\bibfnamefont {A.}~\bibnamefont
  {Tamii}}, \bibinfo {author} {\bibfnamefont {I.}~\bibnamefont {Poltoratska}},
  \bibinfo {author} {\bibfnamefont {P.}~\bibnamefont {von Neumann-Cosel}},
  \bibinfo {author} {\bibfnamefont {Y.}~\bibnamefont {Fujita}}, \bibinfo
  {author} {\bibfnamefont {T.}~\bibnamefont {Adachi}}, \bibinfo {author}
  {\bibfnamefont {C.~A.}\ \bibnamefont {Bertulani}}, \bibinfo {author}
  {\bibfnamefont {J.}~\bibnamefont {Carter}},  \emph {et~al.},\ }\href
  {\doibase 10.1103/PhysRevLett.107.062502} {\bibfield  {journal} {\bibinfo
  {journal} {Phys. Rev. Lett.}\ }\textbf {\bibinfo {volume} {107}},\ \bibinfo
  {pages} {062502} (\bibinfo {year} {2011})}\BibitemShut {NoStop}%
\bibitem [{\citenamefont {Reinhard}\ and\ \citenamefont
  {Nazarewicz}(2010)}]{PhysRevC.81.051303}%
  \BibitemOpen
  \bibfield  {author} {\bibinfo {author} {\bibfnamefont {P.-G.}\ \bibnamefont
  {Reinhard}}\ and\ \bibinfo {author} {\bibfnamefont {W.}~\bibnamefont
  {Nazarewicz}},\ }\href {\doibase 10.1103/PhysRevC.81.051303} {\bibfield
  {journal} {\bibinfo  {journal} {Phys. Rev. C}\ }\textbf {\bibinfo {volume}
  {81}},\ \bibinfo {pages} {051303} (\bibinfo {year} {2010})}\BibitemShut
  {NoStop}%
\bibitem [{\citenamefont {Sarriguren}\ \emph {et~al.}(2007)\citenamefont
  {Sarriguren}, \citenamefont {Gaidarov}, \citenamefont {Guerra},\ and\
  \citenamefont {Antonov}}]{PhysRevC.76.044322}%
  \BibitemOpen
  \bibfield  {author} {\bibinfo {author} {\bibfnamefont {P.}~\bibnamefont
  {Sarriguren}}, \bibinfo {author} {\bibfnamefont {M.~K.}\ \bibnamefont
  {Gaidarov}}, \bibinfo {author} {\bibfnamefont {E.~M.~d.}\ \bibnamefont
  {Guerra}}, \ and\ \bibinfo {author} {\bibfnamefont {A.~N.}\ \bibnamefont
  {Antonov}},\ }\href {\doibase 10.1103/PhysRevC.76.044322} {\bibfield
  {journal} {\bibinfo  {journal} {Phys. Rev. C}\ }\textbf {\bibinfo {volume}
  {76}},\ \bibinfo {pages} {044322} (\bibinfo {year} {2007})}\BibitemShut
  {NoStop}%
\bibitem [{\citenamefont {Angeli}\ and\ \citenamefont
  {Marinova}(2013)}]{ADNDT.99.69}%
  \BibitemOpen
  \bibfield  {author} {\bibinfo {author} {\bibfnamefont {I.}~\bibnamefont
  {Angeli}}\ and\ \bibinfo {author} {\bibfnamefont {K.}~\bibnamefont
  {Marinova}},\ }\href {\doibase https://doi.org/10.1016/j.adt.2011.12.006}
  {\bibfield  {journal} {\bibinfo  {journal} {At. Data Nucl. Data Tables}\
  }\textbf {\bibinfo {volume} {99}},\ \bibinfo {pages} {69} (\bibinfo {year}
  {2013})}\BibitemShut {NoStop}%
\bibitem [{\citenamefont {Krasznahorkay}\ \emph {et~al.}(1999)\citenamefont
  {Krasznahorkay} \emph {et~al.}}]{Krasznahorkay:1999zz}%
  \BibitemOpen
  \bibfield  {author} {\bibinfo {author} {\bibfnamefont {A.}~\bibnamefont
  {Krasznahorkay}} \emph {et~al.},\ }\href {\doibase
  10.1103/PhysRevLett.82.3216} {\bibfield  {journal} {\bibinfo  {journal}
  {Phys. Rev. Lett.}\ }\textbf {\bibinfo {volume} {82}},\ \bibinfo {pages}
  {3216} (\bibinfo {year} {1999})}\BibitemShut {NoStop}%
\bibitem [{\citenamefont {Hashimoto}\ \emph {et~al.}(2015)\citenamefont
  {Hashimoto} \emph {et~al.}}]{Hashimoto:2015ema}%
  \BibitemOpen
  \bibfield  {author} {\bibinfo {author} {\bibfnamefont {T.}~\bibnamefont
  {Hashimoto}} \emph {et~al.},\ }\href {\doibase 10.1103/PhysRevC.92.031305}
  {\bibfield  {journal} {\bibinfo  {journal} {Phys. Rev. C}\ }\textbf {\bibinfo
  {volume} {92}},\ \bibinfo {pages} {031305} (\bibinfo {year} {2015})},\
  \Eprint {http://arxiv.org/abs/1503.08321} {arXiv:1503.08321 [nucl-ex]}
  \BibitemShut {NoStop}%
\bibitem [{\citenamefont {Zenihiro}\ \emph {et~al.}(2018)\citenamefont
  {Zenihiro}, \citenamefont {Sakaguchi}, \citenamefont {Terashima},
  \citenamefont {Uesaka}, \citenamefont {Hagen}, \citenamefont {Itoh},
  \citenamefont {Murakami}, \citenamefont {Nakatsugawa}, \citenamefont
  {Ohnishi}, \citenamefont {Sagawa}, \citenamefont {Takeda}, \citenamefont
  {Uchida}, \citenamefont {Yoshida}, \citenamefont {Yoshida},\ and\
  \citenamefont {Yosoi}}]{zenihiro2018direct}%
  \BibitemOpen
  \bibfield  {author} {\bibinfo {author} {\bibfnamefont {J.}~\bibnamefont
  {Zenihiro}}, \bibinfo {author} {\bibfnamefont {H.}~\bibnamefont {Sakaguchi}},
  \bibinfo {author} {\bibfnamefont {S.}~\bibnamefont {Terashima}}, \bibinfo
  {author} {\bibfnamefont {T.}~\bibnamefont {Uesaka}}, \bibinfo {author}
  {\bibfnamefont {G.}~\bibnamefont {Hagen}}, \bibinfo {author} {\bibfnamefont
  {M.}~\bibnamefont {Itoh}}, \bibinfo {author} {\bibfnamefont {T.}~\bibnamefont
  {Murakami}}, \bibinfo {author} {\bibfnamefont {Y.}~\bibnamefont
  {Nakatsugawa}}, \bibinfo {author} {\bibfnamefont {T.}~\bibnamefont
  {Ohnishi}}, \bibinfo {author} {\bibfnamefont {H.}~\bibnamefont {Sagawa}},
  \bibinfo {author} {\bibfnamefont {H.}~\bibnamefont {Takeda}}, \bibinfo
  {author} {\bibfnamefont {M.}~\bibnamefont {Uchida}}, \bibinfo {author}
  {\bibfnamefont {H.~P.}\ \bibnamefont {Yoshida}}, \bibinfo {author}
  {\bibfnamefont {S.}~\bibnamefont {Yoshida}}, \ and\ \bibinfo {author}
  {\bibfnamefont {M.}~\bibnamefont {Yosoi}},\ }\href@noop {} {\enquote
  {\bibinfo {title} {Direct determination of the neutron skin thicknesses in
  $^{40,48}$ca from proton elastic scattering at $e_p = 295$ mev},}\ }
  (\bibinfo {year} {2018}),\ \Eprint {http://arxiv.org/abs/1810.11796}
  {arXiv:1810.11796 [nucl-ex]} \BibitemShut {NoStop}%
\bibitem [{\citenamefont {Roca-Maza}\ \emph {et~al.}(2011)\citenamefont
  {Roca-Maza}, \citenamefont {Centelles}, \citenamefont {Vinas},\ and\
  \citenamefont {Warda}}]{Roca-Maza:2011qcr}%
  \BibitemOpen
  \bibfield  {author} {\bibinfo {author} {\bibfnamefont {X.}~\bibnamefont
  {Roca-Maza}}, \bibinfo {author} {\bibfnamefont {M.}~\bibnamefont
  {Centelles}}, \bibinfo {author} {\bibfnamefont {X.}~\bibnamefont {Vinas}}, \
  and\ \bibinfo {author} {\bibfnamefont {M.}~\bibnamefont {Warda}},\ }\href
  {\doibase 10.1103/PhysRevLett.106.252501} {\bibfield  {journal} {\bibinfo
  {journal} {Phys. Rev. Lett.}\ }\textbf {\bibinfo {volume} {106}},\ \bibinfo
  {pages} {252501} (\bibinfo {year} {2011})},\ \Eprint
  {http://arxiv.org/abs/1103.1762} {arXiv:1103.1762 [nucl-th]} \BibitemShut
  {NoStop}%
\end{thebibliography}%

\end{document}